# Title

## Maximum Strength and Dislocation Patterning in Multi-principal Element Alloys

**Authors**

Penghui Cao,[1,*]

**Affiliation**

[1]Department of Mechanical and Aerospace Engineering, University of California; Irvine, CA 92697, USA.

*Corresponding author. Email: caoph@uci.edu

**Abstract**

Multi-principal element alloys (MPEAs), commonly termed as medium- or high-entropy alloys containing three or more components in high concentrations, render a tunable chemical short-range order (SRO). Leveraging large-scale atomistic simulations, we probe the limit of Hall-Petch strengthening and deformation mechanisms in a model CrCoNi alloy and unravel chemical short-range ordering effects. It is found that, in the presence of SRO, the maximum strength is appreciably increased, and the strongest grain size drifts to a small value. Additionally, the propensity for faulting and deformation transformation is reduced and accompanied by the intensification of planar slip and strain localization. We reveal strikingly different deformation microstructures and dislocation patterns that prominently depend on crystallographic grain orientation and the number of slip planes activated during deformation. Grain of single planar slip attains the highest volume fraction of deformation-induced structure transformation, and grain with double active slip planes develops the densest dislocation network. These results advancing the fundamental understanding of deformation mechanisms and dislocation patterning in MPEAs suggest a mechanistic strategy for tuning mechanical behavior through simultaneously tailoring grain texture and local chemical order.

## Introduction

The strength and hardness of polycrystalline metals increase with reducing their grain size, as a high stress is required for nucleation and glide of dislocations in a confined volume. The grain refinement strengthening, understood as Hall-Petch effect (*1, 2*), can impart ultra-high strength to materials; for example, the strength of copper exhibits nearly two-order magnitude enhancement and achieves a gigapascal strength with decreasing the crystallite size from micrometer to nanometer scale (*3, 4*). However, as the grain size falls into a dozen nanometers, grain boundary strengthening breaks down and softening takes over as the predominate mechanism, giving rise to the strongest size (*5, 6*) at which the material reaches its maximum strength. The classic Hall-Petch scaling, originally proposed for pure metals, has been proved to be valid (*7, 8*) in an emergent class of multi-principal element alloys (MPEAs). Comprising multiple principal components in high concentrations, the MPEAs, including medium- and high-entropy alloys (M/HEAs) are presumed to be random solid solutions corresponding to maximum configurational entropy (*9*). The ideal random solid solutions in MPEAs, however, may only be possible at temperatures close to melting point (*10*). As the temperature decreases, solute-solute interaction and mixing enthalpy (enthalpic contribution) predominate the Gibbs free energy and induce local ordering of the chemistry. Indeed, appreciable short-range order (SRO) has emerged in long-time annealed MPEAs, and considerably impacts dislocation slip and mechanical behavior (*11–13*). As a salient feature pertaining to MPEAs, an intriguing question remains as to whether or how SRO affects the strongest size, maximum strength, and the underlying deformation mechanisms.

Within the numerous MPEA systems, the face-centered cubic (*fcc*) CrCoNi-based alloys have attracted extensive attention owing to their extraordinary mechanical properties, including high strength and ductility (*8, 10*). The mechanistic origin of superior mechanical performance is traced to their low stacking faculty energy, which facilitates twining and phase transformation during mechanical straining. Twinning-induced plasticity and transformation-induced plasticity, known as TWIP and TRIP effects from advanced steels (*14*), provide extra interface hardening and defer the onset of necking, contributing to an enhanced strength-ductility synergy. When the SRO is engineered into the materials, both stacking fault energy and activation energy barrier of dislocation motion are increased (*15*), which likely alters dislocation slip pathway and hence deformation microstructures (for instance, nanotwin and deformation transformation). Besides the intrinsic material properties, another parameter that would control deformation microstructure of grain is its crystallographic orientation, because it determines the forms of active slip system, e.g., single planar, double planar, and multiple planar slip (*16*). While planar slip bands, deformation nanotwins, and deformation hexagonal close-packed (*hcp*) structures are widely observed in experiments (*17–19*), understanding the underpinning dislocation mechanism and their relationship with the grain orientation remains a challenging problem due to the inability of in-situ tracking dislocations in three-dimensional bulk materials during straining. Leveraging on large-scale deformation simulations of a model *fcc* CrCoNi alloy at the atomistic level, we uncover strikingly different characteristics of dislocation patterning and deformation microstructure evolution, substantially depending on grain orientation.

## Results

Large-scale polycrystal models consisting of grain size from 40 to 3 nm and involving up to 97.3 million atoms are adopted. We use hybrid Monte Carlo (MC) and molecular dynamics (MD) simulation to obtain equilibrated systems with SRO (see Methods);

hereafter, we refer SRO and RSS as polycrystalline systems with short-range order and random solid solution, respectively. Figures 1a-d show the structure and chemical mapping of an annealed polycrystal with grain size of 32 nm. Detailed analysis on chemical distribution and local order in all models confirms the same trend of chemical SRO in grains and grain boundaries (Supplementary Figures 1-2). Subjected to uniaxial tensile deformation, the stress-strain responses of polycrystals display an initial elastic regime followed by plastic flow (Supplementary Figure 3). In Figure 1e, we demonstrate the variation of plastic flow strength of polycrystals with their constituent grain size. The grain size dependence of strength clearly reveals that the maximum strength of SRO systems is appreciably raised as compared to RSS. The critical grain size (strongest size) corresponding to the highest strength drifts to a smaller value in SRO. This suggests that the introduced chemical short-range order strengthens the materials, stabilizes grain boundaries, and extends the Hall-Petch effect to smaller nanograins.

To understand the effects of local chemical order on deformation mechanisms, Figs. 2a and b show the typical deformation microstructures at 20% strain for RSS and SRO samples, respectively. A high fraction of *hcp*-coordinated atoms appear in the *fcc* grain and manifest in the form of stacking fault (SF), *hcp* phase, and twin boundary (TB). The morphology of these *hcp*-type structures varies from one grain to another, clearly indicating grain orientation dependence of deformation substructures (to be discussed later). In contrast with RSS, the SRO sample exhibits a reduced fraction of *hcp* atoms. That can be seen in Fig. 2c, where the volume fraction of deformation-induced *hcp* atoms is plotted as a function of tensile strain. The variation of *hcp* fraction with strain demonstrates three-stage behaviors, referred to as incubation period in elastic deformation regime, followed by rapid growth after yielding, and lastly, the steady and slow growth during plastic flow stage. At 20% tensile strain, the SRO system accumulates 33 volume percent of *hcp* structure, which is nearly 15% lower than that of RSS. The decrease of *hcp* atom fraction in SRO is closely related to SRO-induced SFE increase (*15*, *20*) and glide plane softening(*21*), as the former lowers the propensity for nucleation and growth of *hcp* lamellae. The latter resulting from local SRO destruction promotes repetitive partial dislocation slip in the slip-softened layers such as SF and *hcp* lamella. Since no footprint (defects) will be left over in the grains after successive passage of leading partial and trailing partial, the instantaneous deformation structure is impotent to trace deformation history the sample has been experienced. To evaluate plastic deformation strain localization, we calculate the atomic-level plastic strain (*22*) $\eta$ and its spatial distributions are shown in the right panels of Figs. 2a and b. In comparing the strain maps, one can see that SRO exhibits extended and strain-concentrated planar slip nano-bands, implying repetitive dislocation glide takes place in the band regions. The increased strain localization, originating from SRO destruction and glide plane softening, is quantitatively depicted in Fig. 2d, in which the statistical distributions of large local strain ($\eta > 1$) are shown. From extreme value theory aspect (*23*), the SRO material carrying large plastic strain bursts (the long tail in the $\eta$ distribution) is prone to cause uncoordinated deformation and initiate decohesion events (nano-crack formation).

Governing by grain orientation and Schmid's law, a different number of slip planes can be activated under mechanical straining, and substantially influence, if not redetermine, dislocation patterns and deformation microstructures. Figure 3 presents a deformed grain with a single primary slip plane activated, where deformation microstructure, local strain map, and dislocation configurations are delineated. The smoothly curved and long Shockley partial dislocations, emitted from grain boundary, glide on parallel (111) planes (i.e., triangle ABC in Thompson tetrahedron) without intersection (Fig. 3e). The resultant local strain is uniformly spreading crossing these planes (Fig. 3b). Arising from partial dislocation glide, SF is produced behind a moving dislocation in *fcc* structure. Incipient *hcp* phase is nucleated when two partials glide on every other (111) plane. The deformation microstructure (Fig. 3a) exhibits piles of *hcp* lamella and SF (including intrinsic and extrinsic SF) as detailed by Fig. 3c. These deformation-induced *hcp* structures impede dislocations on other intersecting slip planes (benefiting from grain rotation during deformation), confining them in the grain corners (Fig. 3e). In Fig. 3d, we present the strain variation of the volume fraction of all *hcp*-coordinated atoms (red-colored circles), SF (purple), *hcp* phase (green), and TB (yellow). In the early stage of plastic deformation (5% strain), the *hcp* structures predominantly are SFs, and they grow rapidly with strain. From 10% strain, *hcp* phase growth, i.e., martensitic transformation (*fcc* to *hcp*), turns into the primary deformation mechanism, which consumes the existing SFs (decrease of SF in Fig. 3d). At 20% strain, 45 volume percent of the *fcc* phase is transformed to *hcp* structures manifested as hcp phase/lamella (28%), SFs (16%), and TB (1%) (Supplementary Video 1 for microstructure and dislocation evolution).

The grain having two active slip planes exhibits distinct deformation microstructure and dislocation patterns (Fig. 4). Dislocations that glide on two intersecting planes of (111) and ($\bar{1}1\bar{1}$) (i.e., ABC and BCD in Thompson's notation) collide into each other. Their reaction generates a high density of sessile dislocations: stair-rod partial and Hirth partial (Fig. 4e), depending on the direction of Burgers vectors (*24*). Owing to dislocation slip, the grain is gridded and develops an *hcp*-structure network (Fig. 4a). It is noteworthy that TBs are always associated with intersection lines (Fig. 4c), implying that twinning dislocation is mediated by collision and reaction. Compared with the grain of a single planar slip, the current grain features a faster accumulation of *hcp* atoms right after yielding (5% strain in Fig. 4d) because of simultaneous slip on two planar systems. But during plastic flow stage, the growth of hcp-structure is slower and reaches 32% volume fraction at 20% strain (structure volume and dislocation density summarized in Supplementary Table 1). Concerning individual *hcp*-type structures, SFs become saturated at 10% strain. In the meantime, both *hcp* phase and TB begin to progress rapidly, leading to matristic transformation and twinning deformation in plastic flow (Supplementary Video 2).

Figure 5 shows the deformation microstructure and dislocation configurations in a grain of multiple slip planes activated in deformation. From structure aspect, the deformation-induced *hcp* slices, lying in their own slip planes, intersect and form three-dimensional networks (Figs. 5 a and c). Examining microstructure evolution (Fig. 5d), we can see that after yielding, a large fraction of *hcp* structures is formed but immediately becomes saturated during plastic flow. The entire *hcp* structures, which are mainly SFs, occupy less than 20% of the grain volume at the end of deformation. Intriguingly, twinning deformation is unfavored when multiplanar slippage is activated, and matristic transformation is appreciably suppressed, as indicated by the reduction of *hcp* phase in the plastic flow stage (7-10% strain in Fig. 5d). As a result of dislocation motion on multiple intersecting planes, a dense dislocation line network comprised of glissile and sessile dislocations is developed (Fig. 5c). These sessile stair-rod dislocations and Hirth partials bridge Shockley partials on various slip planes, presumably acting as the backbone of dislocation network (Supplementary Video 3).

**Discussion**

At a given temperature and strain rate, the strength of MPEAs is dictated by dislocation motion-related solid solution strengthening and grain size dependent Hall-Petch effect. In the presence of SRO and consequently high SFE (*15*, *20*), it is reasonable to speculate the motion of leading partial will experience a large resistance and high energy barrier. Additionally, the following passage of trailing partial, when it happens, can erase the SF and restore *fcc* structure, but inevitably create an anti-phase boundary owing to destruction of SRO on the slip plane. The interplay between stacking fault creation, anti-phase boundary generation, and the associated energy penalty, stemming from the existence of SRO, exerts back stress on moving dislocation and retards its motion, which is the root cause of enhanced solid-solution strengthening. The Hall-Petch strengthening is influenced by SRO in a way that the critical grain size (strongest size) corresponding to the maximum strength changes to a smaller value, accompanied by deferred strength softening (Fig. 1e). This implies grain boundary-mediated plasticity, such as boundary sliding and migration responsible for the softening(*6*), is alleviated by the introduced SRO. The mitigation of boundary softening likely benefits from SRO-lowered boundary enthalpy that improves its sliding resistance. The SRO in the grain interior would also hinder grain boundary migration because, when a grain boundary sweeps its grain matrix, the chemical SRO in the swept volume will be destroyed as the boundary migration alters the crystallographic orientation of the swept matrix(*25*). The grain size-dependent chemical ordering in grain boundary and grain matrix stabilities boundary network and mitigates softening effects, deserving careful attention and further study. It is worth noting that the imposed straining rate ($10^8 \ s^{-1}$) in the simulations, which can adequately enable true dislocation dynamics (*26*), is still several orders of magnitude higher than that of conventional laboratory experiments. When sufficiently lowering the strain rate, thermally activated diffusional processes, including vacancy-mediated dislocation climb and Coble creep, can take place and decrease the strength. Because chemical SRO considerably reduces and localizes diffusion (*27*), it is tempting to hypothesize that the strength difference between RSS and SRO systems will widen even further at reduced strain rates and/or elevated temperatures.

Essentially rooted in dislocation motion, deformation microstructure evolution and dislocation pattern formation have remained as the central focus of physical metallurgy in order to interpret and ultimately predict mechanical behavior. The ultra-large atomistic simulations, capturing the individual atomic motions, enable the study of dislocation motion and patterning at the fundamental level. Especially it elucidates the atomistic processes of deformation twinning and martensitic transformation—the deformation mechanisms the other mesoscale simulation methods (*28*) like discrete dislocation dynamics are inadequate to conquer. Owing to the SRO imparted to the system and resulting high SFE and anti-phase boundary, different deformation processes, including dislocation slip, faulting, twinning, and phase transformation, are influenced and balanced so that system deforms along its minim energy pathways. We found that, in the existence of local chemical order, faulting and deformation transformation are suppressed and planar slip is increased (glide plane softening and repetitive slip), which gives rise to a decrease of *hcp* structure density and exacerbation of strain localization in SRO alloys.

The deformation microstructure and dislocation density highly depend on crystallographic grain orientation ( Table 1). With a single active slip plane, grain accommodates the highest density of deformation *hcp* lamella, resulting from partial dislocation glide on every other plane. The majority of dislocations are glissile Shockey partials, gliding on parallel planes and scarcely colliding into each other. For the grain of double planar slip, dislocation moving on the intersecting planes results in collision and reaction, leading to formation of Hirth partials and stair-rod dislocations. A significant number of deformation nanotwins appear in the deformed grain, signifying that twin formation and growth are closely related to dislocation interactions. When multiple planes are triggered, the volume density of deformation-induced *hcp* atoms evolves into the lowest. This reduced propensity for *hcp* accumulation originates from multiplanar dislocation interactions, which produce various partial dislocations and can cause both *hcp* layer thickening (*fcc* to *hcp*) and thinning (*hcp* to *fcc*), depending on the Burgers vector. A dense and intertwined dislocation network, consisting of sessile dislocations and interspersed Shockley partials, emerges from an increased probability of dislocation collision.

In summary, the study reveals the maximum strength and critical grain size can be manipulated by tuning the degree of chemical SRO, signifying the non-trivial role of local chemical ordering on achieving ultimate strength of MPEAs. As the salient feature of MPEAs, the presence of SRO considerably impacts deformation microstructure and local plastic strain, originating from its effects on stacking fault energy increase and anti-phase boundary development. As a result, the volume fracture of deformation *hcp* atoms is decreased to compensate the increased stacking fault energy, and plastic strain is more localized to embrace slip-mediated SRO destruction and glide plane softening. Most importantly, the crystal orientation of constituent grains plays a vital role in deformation mechanisms, microstructure evolution, and dislocation patterning, which collectively impact the mechanical behavior of individual grains and hence system-level performance. Moving forward, the emerging MPEAs providing vast compositional space that remains to be explored enable the tunable chemical order at short- or even medium-range distances, which together with tailorable grain texture, offer a promising avenue for attaining controllable deformation mechanisms and extraordinary mechanical response beyond the traditional dilute alloys.

**Materials and Methods**

   **Molecular dynamics simulation.** We create two independent polycrystal models, respectively consisting of 8 and 27 grains, to study the grain size effects on strength. To construct different grain-sized systems, the crystallographic grain orientations in the polycrystal are kept the same when scaling the structure and grain to various sizes. Total 40 samples are created and simulated that contain 0.67 to 97.32 million atoms and cover grain size range of 3 to 40 nm. We use a model embedded-atom method potential(*15*) for CrCoNi alloy, and it enables two control systems, RSS and SRO, for studying the effects of the presence of SRO (importantly stacking fault energy variation and anti-phase boundary generation) on deformation mechanisms. With periodic boundary conditions imposed in all three directions, uniaxial tensile deformation is applied to the samples at a constant engineering strain rate of $10^8$ s$^{-1}$ and temperature of 300 K using Nosé–Hoover thermostat. The mechanical stresses of other directions perpendicular to tension are controlled at zero using Parrinello-Rahman barostat. The atomistic simulations of deformation are carried out using open-source code LAMMPS(*29*), and atomic structure visualization and dislocation representation are rendered with OVITO(*30*) and dislocation extraction algorithm(*31*).

   **Monte Carlo and molecular dynamics simulation, and short-range order.** The hybrid Monte Carlo (MC) and molecular dynamics (MD) simulations within the variance-constrained semi-grand-canonical (VC-SGC) ensemble(32), are performed to construct the systems with chemical SRO. A total of 500,000 MD timesteps with a step size of 1 fs are carried out, and for every 50 timesteps the MD simulation is interrupted to execute MC cycle. In the MC simulations, we carry out $N/3$ number of atom type swap trails, where $N$ is the total atoms in the system. In the end of simulation, each atom in average has subjected to 3,300 atom type change trials. The acceptance probability is determined by five parameters(32), including system energy change $\Delta u$ after a trail, concertation difference $\Delta c$ from the targeted equimolar concertation, chemical potential difference between two species $\Delta \mu$, variance parameter $k$, and system temperature $T$. We use the parameters $\Delta \mu_{Ni-Co} = 0.021$ and $\Delta \mu_{Ni-Cr} = -0.31$ eV determined from semi-grand canonical ensemble simulation(15), $k = 1,000$, and $T = 650$ K, which result in the equimolar concentration in the end of simulation (<0.085 atom% deviation). In the MD intervals, the system stresses are relaxed to zero average value using Parrinello-Rahman barostat. It is worth noting that the hybrid MC and MD simulation, incapable of modeling the diffusion kinetics associated with SRO formation as in actual aging experiments, produces an equilibrated structure determined by thermodynamics. Nevertheless, it provides a neat low energy state with SRO, which makes it possible to study local chemical effects on deformation mechanisms.

**Short-range order parameters.** To quantify chemical SRO in polycrystals, we modify the non-proportional number(20). The order parameter in the first nearest neighbor shell between any pair of atoms $i$ and $j$ is defined as, $\delta_{ij} = N_{ij} - Z_0 C_j$, where $N_{ij}$ is denotes the actual number of pairs, $Z_0$ represents the coordination number, and $C_j$ is $j$ atom concentration. A positive $\delta_{ij}$ indicates a favored and increased number of pairs, meaning element $i$ tends to bond with element $j$, while a negative value represents an unfavored pairing. For grain atoms, the value of $Z_0$ is 12 (fcc structure), and $Z_0$ has a slightly smaller value of 11.42 for grain boundary atoms. The calculated order parameters of grains and grain boundary at the first nearest shell are presented in Supplementary Fig. 2.

**Deformation SF, hcp phase, and twin boundary identification.** Dislocation slip and plastic deformation of *fcc* materials can introduce a variety of *hcp*-coordinated atomic structures, manifested in the form of intrinsic SF, extrinsic SF, twin boundary, and *hcp* phase (lamella). The most frequently used structure characterization methods, such as common neighbor analysis (CAN)(33, 34) and centro-symmetry parameter analysis(35), which can effectively classify local structure type (for instance, *hcp*) of each atom, are incompetent at differentiating different *hcp* structures. To this end, we define a weighted coordination number $Z$, measuring the number of same structure neighbors an atom has within a cutoff distance $r_c$. When the cutoff distance is covering the two consecutive habit planes (set to be middle of the second (111) and third (111) planes, i.e., $2.5a/\sqrt{3}$, here $a$ is lattice constant), all the hcp structures, including intrinsic SF (iSF), extrinsic SF (eSF), TB, and *hcp* phase can be successfully identified from $Z$ ($Z_{TB} = 18, Z_{eSF} = 24, Z_{iSF} = 30, Z_{hcp} \geq 37$), as shown in Supplementary Fig. 4. The value of $Z$ is calculated via coordination analysis of atoms within $r_c$ and with *hcp* structure, $f(r_c = 2.5a/\sqrt{3}, \text{hcp})$, and well captures the unique feature associated with each defect, realizing the structure identification.

**Figures and Tables**

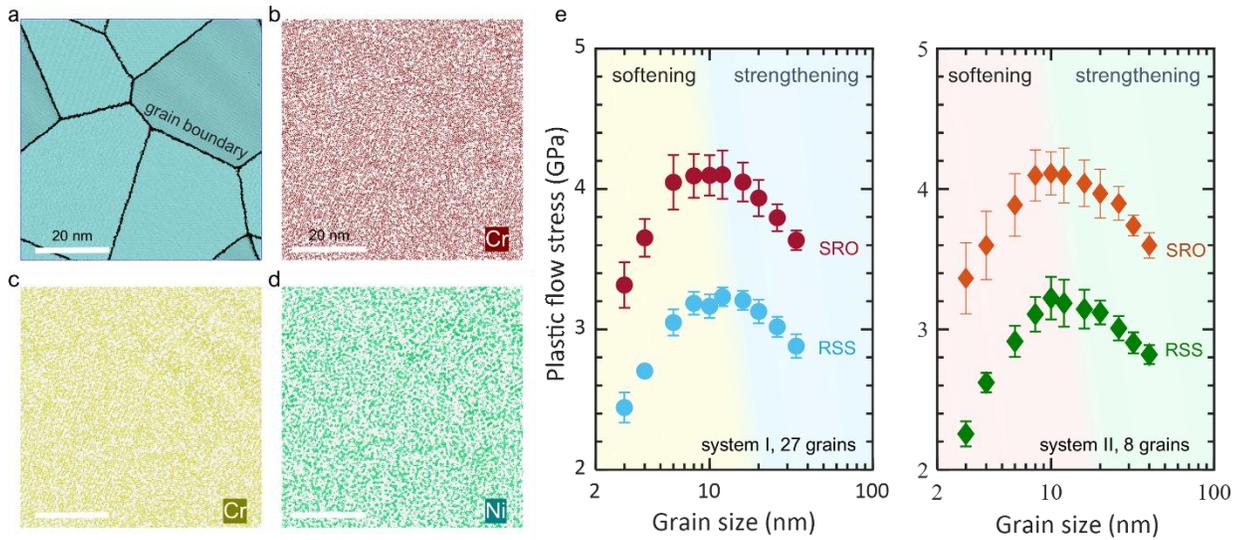

**Fig. 1. Microstructures and Hall-Petch strengthening in CrCoNi alloys. a-d**, Snapshots of polycrystalline microstructure and the corresponding mapping of constituent elements from MC-MD annealing simulations. The detailed results of SRO analysis and element distributions in grains and grain boundaries are summarized in Supplementary Figs. 1-2. **e**, Grain size dependence of plastic flow strength for CrCoNi alloys with RSS and SRO. The two panels correspond to two independent polycrystalline structures consisting of 27 and 8 grains, respectively.

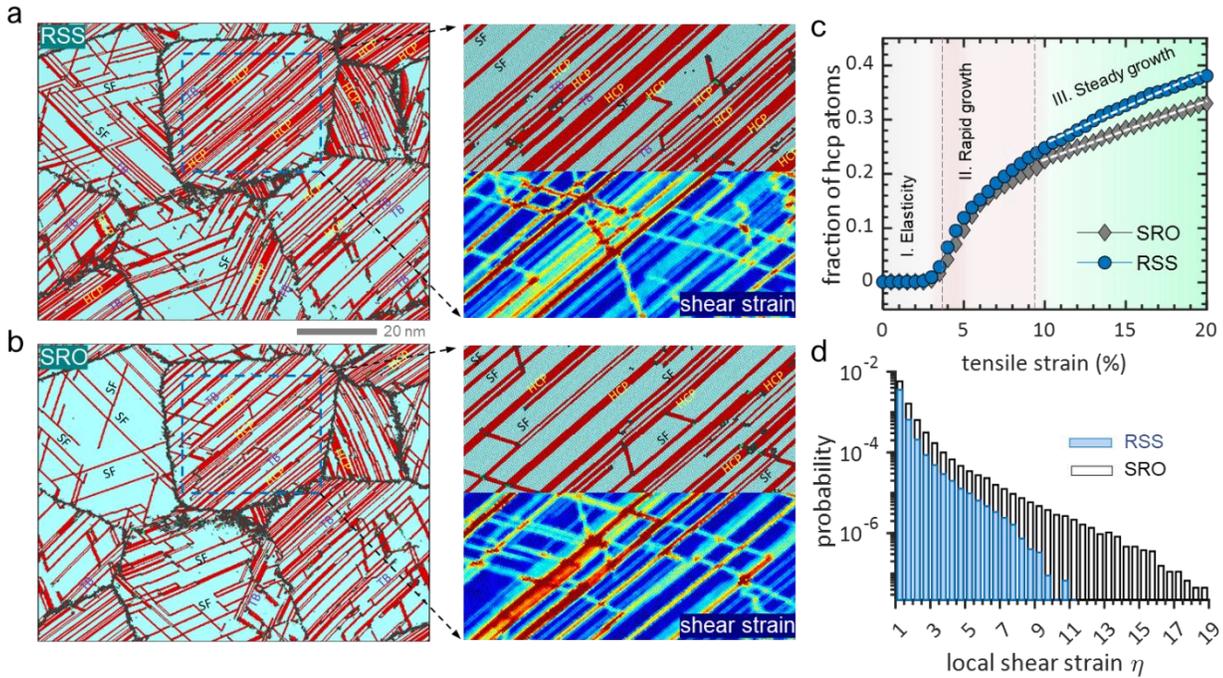

**Fig. 2. Deformation microstructure and local plastic strain. a,b**, Atomic structures and local strain for RSS and SRO samples stretched to 20% strain, respectively. Atoms are color-coded by structural type, where red color represents *hcp*-coordinated atoms, and blue indicates *fcc* structure. **c**, Volume fraction of *hcp*-coordinated atoms as a function of tensile strain. **d**, Probability distributions of local strain in RSS and SRO. The SRO system exhibits long tail and extreme local strain under deformation.

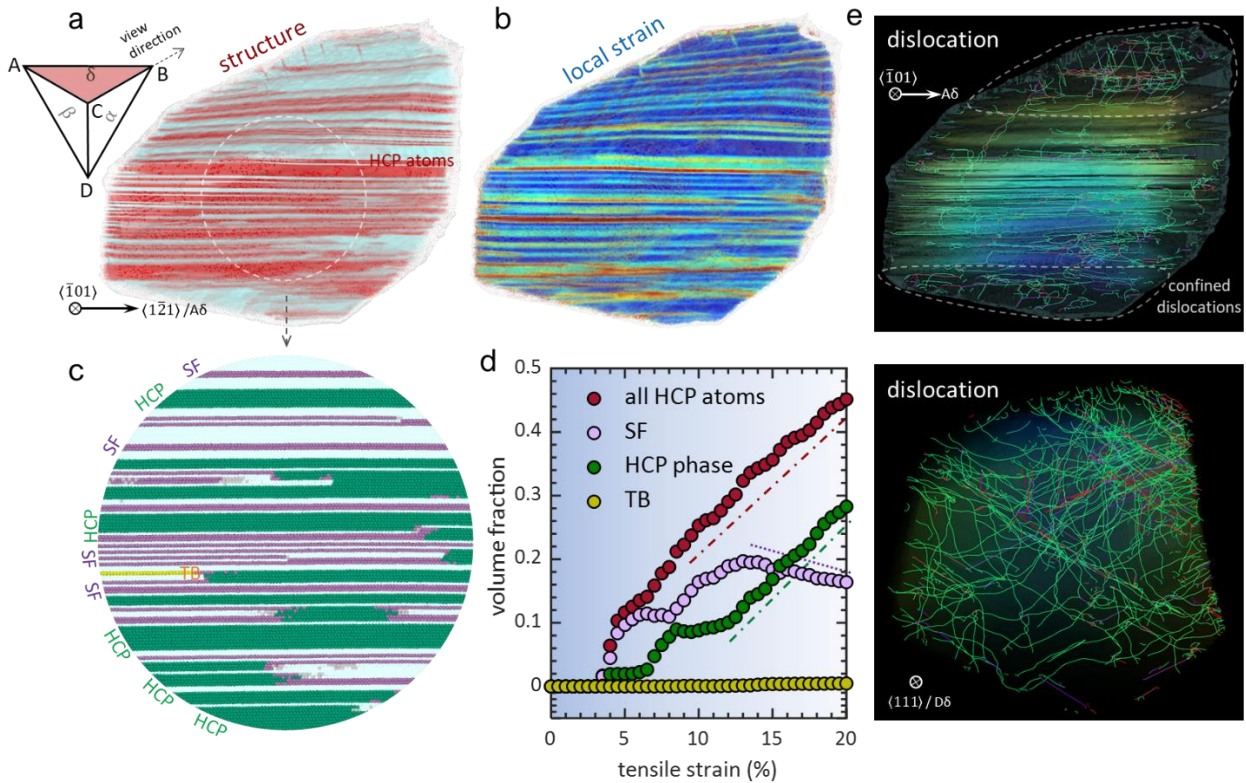

**Fig. 3. Deformation characteristics and dislocation pattern in grain of a single active slip plane. a,b**, Deformation microstructure and local strain map of grain deformed at 20% strain, respectively. Thompson tetrahedron is shown to illustrate active slip plane $ABC$ and view direction $CB$. The hcp-coordinated atoms are colored in red. **c**, Magnification of deformation microstructure in the circled region of (**a**). The *hcp*-coordinated atoms are identified as SF, hcp phase, and TB. **d**, Volume fraction of deformation structures as a function of strain for the grain with single planar slip. **e**, Dislocation configurations in the deformed grain. Smoothly curved Shockley partial dislocations (green) lie in the parallel $ABC$ planes, and stair-rod (purple) and Hirth partial (red) dislocations are scattered.

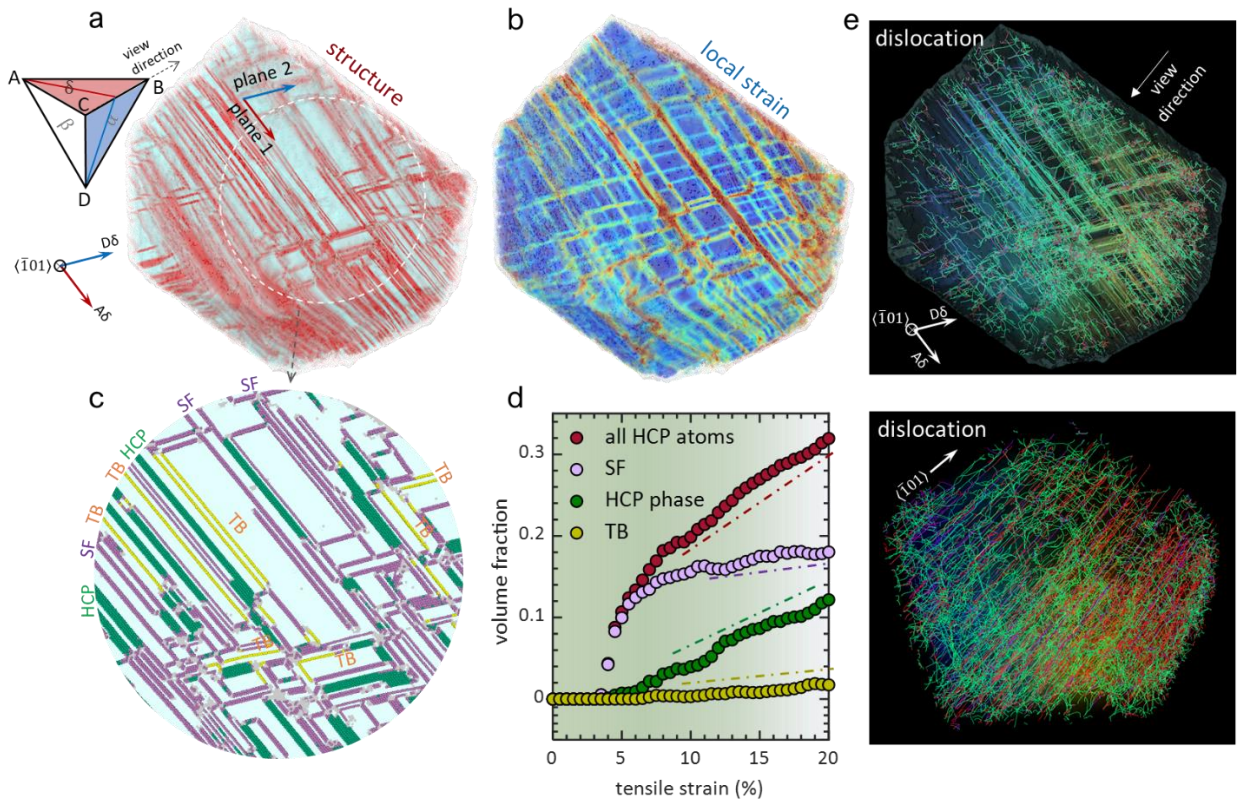

**Fig. 4. Deformation characteristics and dislocation pattern in grain of double active slip planes. a,b**, Deformation microstructure and local strain map of grain at 20% strain, respectively. Thompson tetrahedron is shown to illustrate active slip planes $ABC$ and $BCD$. **c**, Magnification of deformation microstructure in the circled region of (**a**). The *hcp*-coordinated atoms are characterized as SF, hcp phase, and TB. **d**, Volume fraction of deformation structures as a function of strain for the grain with double planar slip. **e**, Dense dislocation network in the deformed grain. Highly curved Shockley partial dislocations (green) lie in two intersecting slip planes, and dense stair-rod (purple) and Hirth partial (red) dislocations are formed in the grain interior.

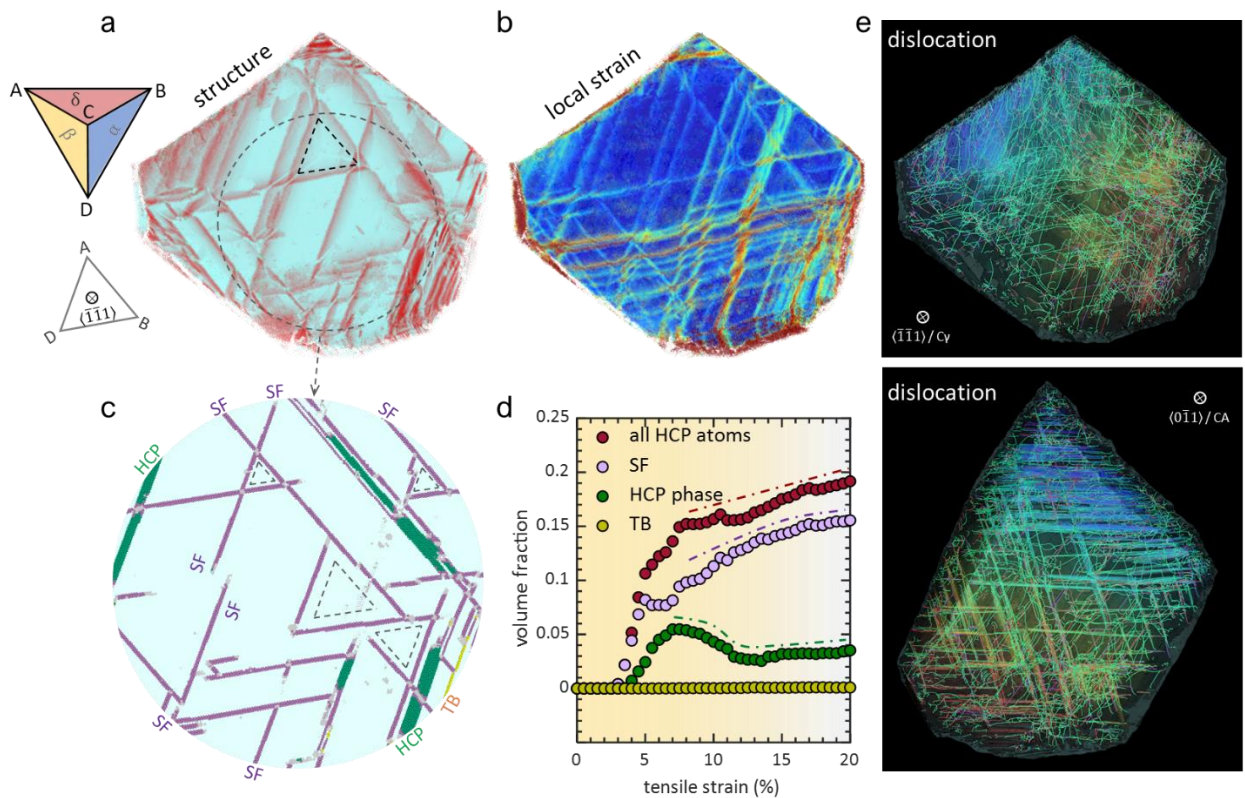

**Fig. 5. Deformation characteristics and dislocation pattern in grain of multiple active slip planes. a,b,** Deformation microstructure and local strain map of grain at 20% strain, respectively. **c,** Magnification of deformation microstructure in the circled region of (**a**). The *hcp*-coordinated atoms are mainly SFs. **d,** Volume fraction of deformation structures as a function of strain for the grain with multiplanar slip. **e,** Dislocation network in the deformed grain shows some local regions with concentrated dislocations. Shockley partial dislocations (green) lie in multiple slip planes, with variously arranged stair-rod (purple) and Hirth partial (red) dislocations.

**Table 1.** Quantitative characterizations of deformation structure and dislocation at 20% tensile strain

| | Deformation structures (volume fraction) | | | | Dislocation statistics (density, /m$^2$) | | | |
|---|---|---|---|---|---|---|---|---|
| | Total hcp atoms | SF | hcp phase | TB | Total density | Shockley | Stair-rod | Hirth partial |
| **Grain 1** one active slip plane | 45.2% | 16.4% | 28.3% | 0.5% | $6.66 \times 10^{16}$ | $5.36 \times 10^{16}$ | $0.45 \times 10^{16}$ | $3.80 \times 10^{16}$ |
| **Grain 2** two active slip planes | 31.9% | 18% | 12.2% | 1.74% | $2.23 \times 10^{17}$ | $1.64 \times 10^{17}$ | $0.23 \times 10^{17}$ | $0.33 \times 10^{17}$ |
| **Grain 3** multiple active slip planes | 19.2% | 15.6% | 3.5% | 0.1% | $1.91 \times 10^{17}$ | $1.30 \times 10^{17}$ | $0.14 \times 10^{17}$ | $0.145 \times 10^{17}$ |

**Table 1.** Statistics of deformation structure and dislocation density in grains with one active slip plane, two slip planes, and multiple slip planes.

**Supplementary Materials**

This manuscript includes Supplementary Materials.

# Supplementary Materials for

## Maximum Strength and Dislocation Patterning in Multi-principal Element Alloys

Penghui Cao

*Corresponding author. Email: caoph@uci.edu

**This PDF file includes:**

Figs. S1 to S4
Captions for Movies S1 to S3

**Other Supplementary Materials for this manuscript include the following:**

Movies S1 to S3

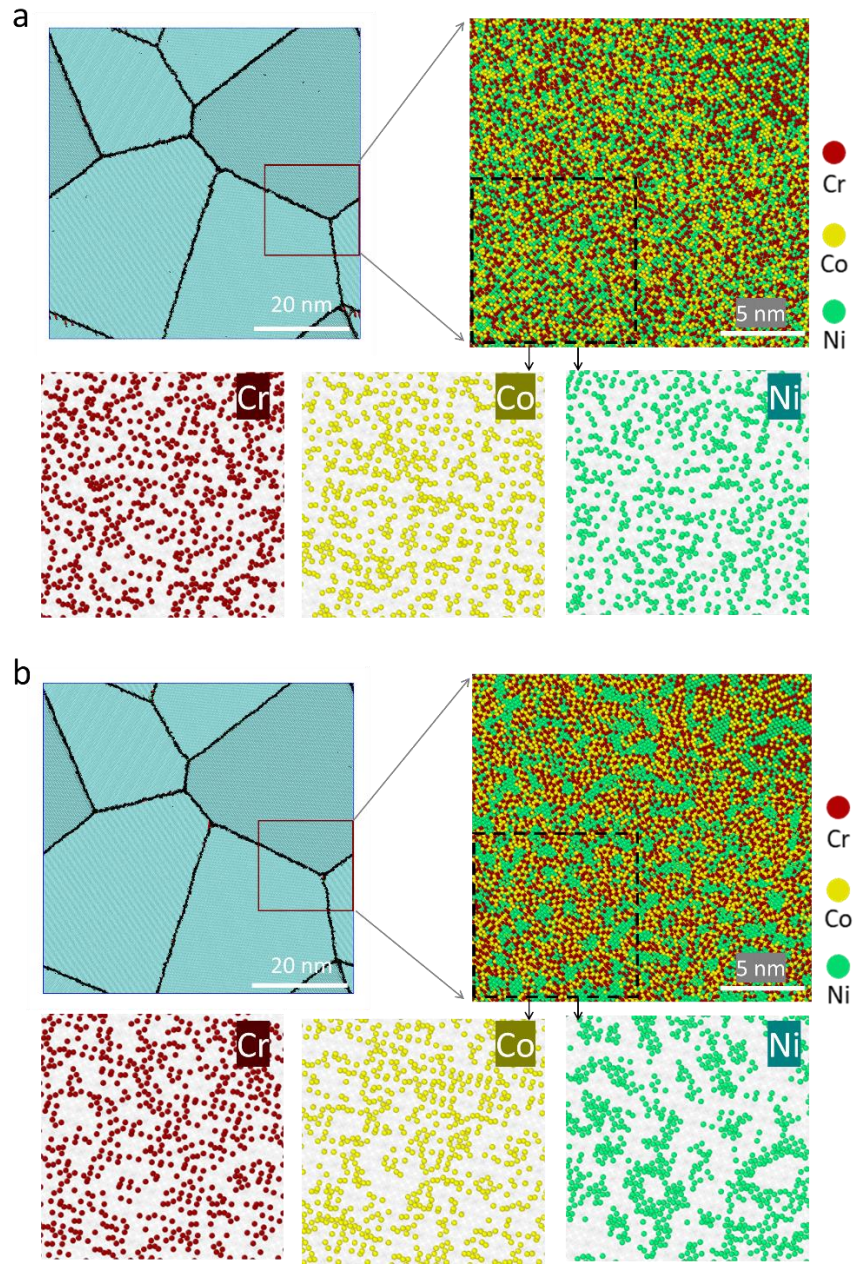

**Fig. S1.**

Structures and element distributions in RSS (a) and SRO (b) CrCoNi alloys. The bottom panel in each figure, corresponding to the magnification of the local region (black dotted box), shows the concentration distribution of individual elements, Cr, Co, and Ni.

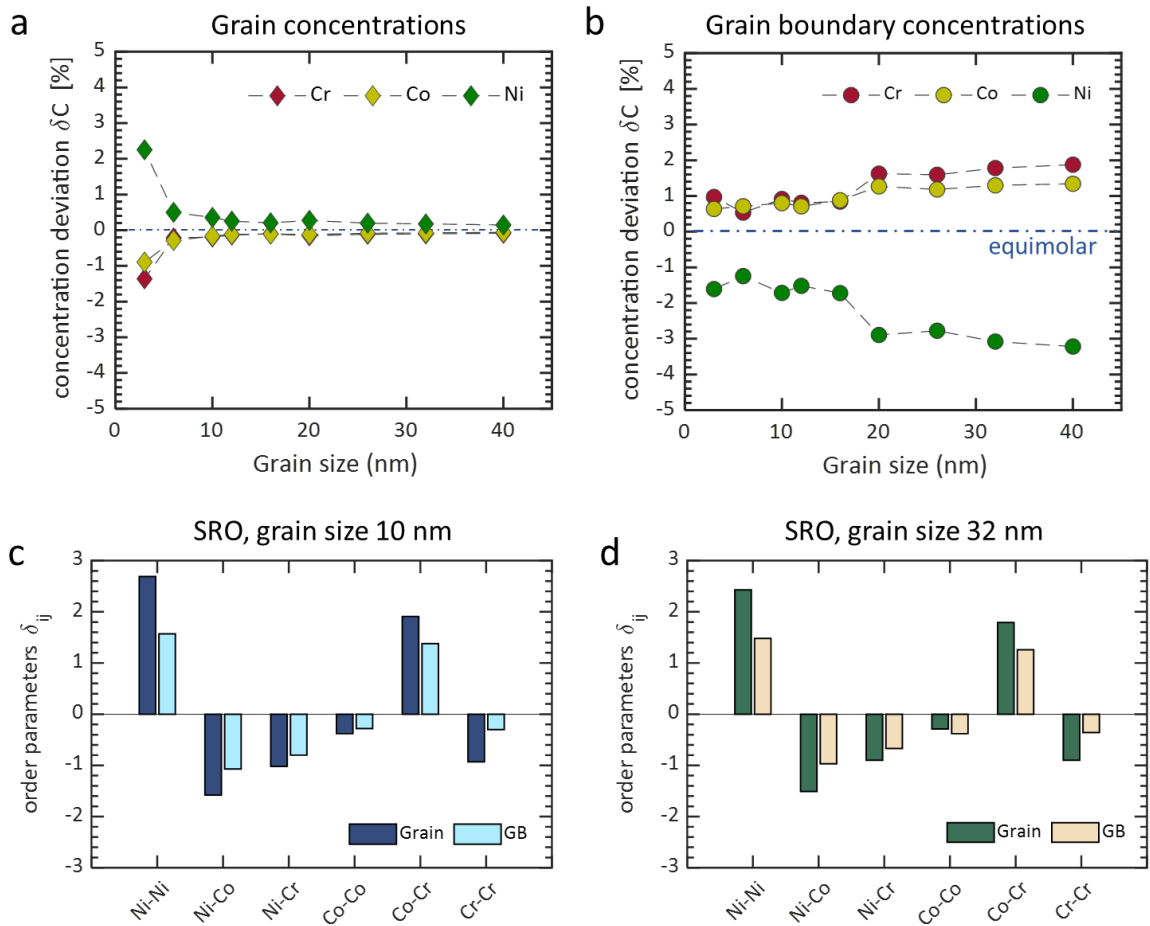

**Fig. S2.**
Concentration variation and chemical short-range order in annealed polycrystalline alloys. a, Grain concentration for polycrystals with grain size from 40 nm to 3 nm. b, Grain boundary concentration for different grain-sized polycrystals. c,d, Short-range order parameters between a pair of atoms $i$ and $j$, $\boldsymbol{\delta_{ij} = N_{ij} - Z_0 C_j}$, where $N_{ij}$ is denotes the actual number of pairs, $Z_0$ represents the coordination number, and $C_j$ is $j$ atom concentration. The order parameters in grain interiors and grain boundaries are evaluated for annealed polycrystals with the grain size of 10 nm (c) and 32 nm (d).

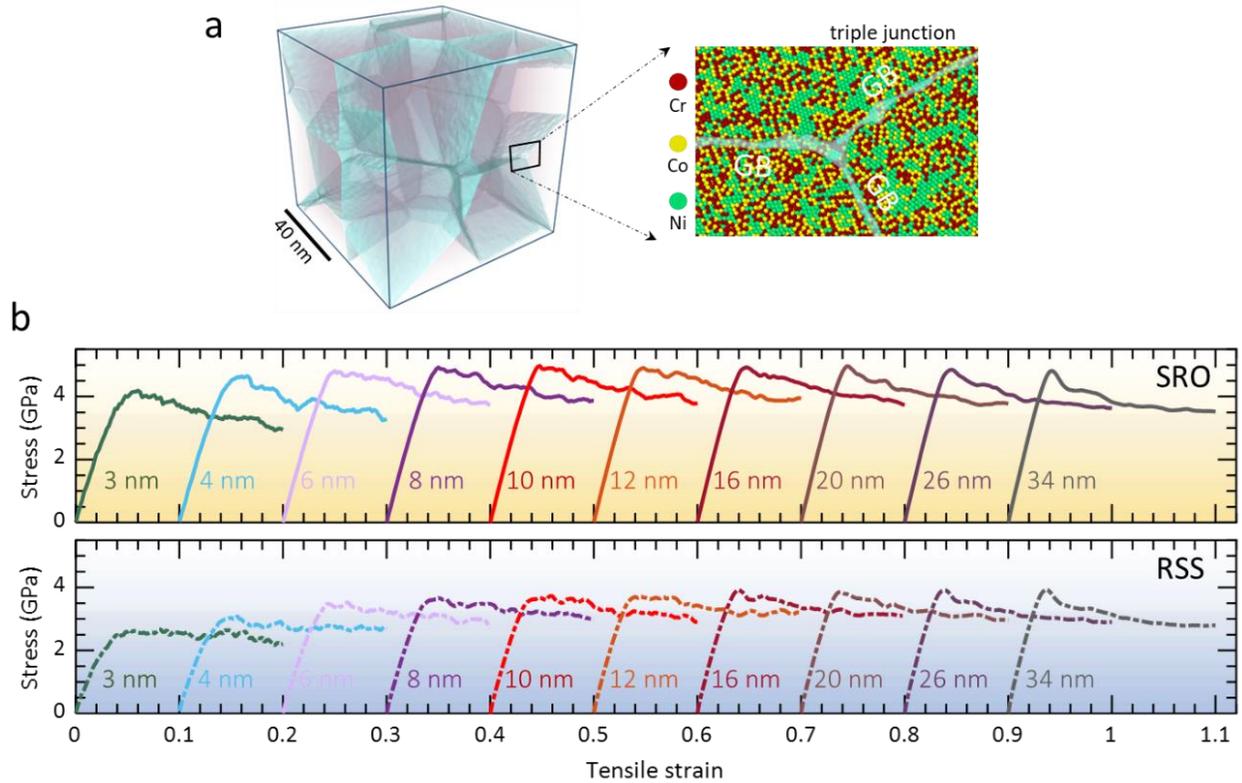

**Fig. S3.**

Mechanical responses of SRO and RSS systems. a, Thermall-dimensional microstructure of the polycrystalline model and local concentration distributions. b, Stress-strain responses of polycrystals subjected to uniaxial tension. The RSS and SRO denote systems of random solid solution and short-range order, respectively. When comparing these strain-strain curves consisting of an initial elastic regime followed by plastic flow, the flow stress gradually shifts upward with decreasing grain size, but, from about 10 nm, shows an opposite trend with continuous reduction of grain size to 3 nm.

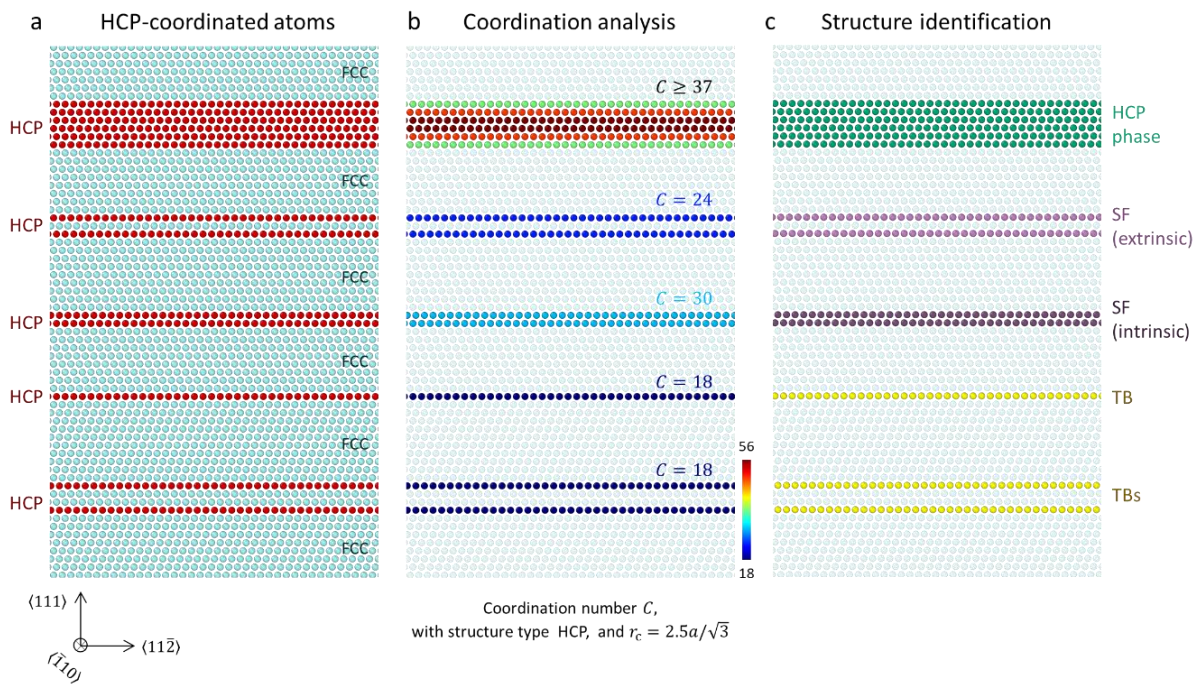

**Fig. S4.**
Deformation SF, hcp phase, and twin boundary identification. a, A *fcc* structure containing hcp-coordinated atoms. b, Atoms are colored by weighted coordination number $Z$, that measures the number of same *hcp*-structure neighbors an atom has within a cutoff distance $2.5a/\sqrt{3}$ (*here $a$* is the lattice constant). c, All the hcp structures, including intrinsic SF (iSF), extrinsic SF (eSF), TB, and *hcp* phase are identified from $Z$ ($Z_{\text{TB}} = 18, Z_{e\text{SF}} = 24, Z_{i\text{SF}} = 30, Z_{\text{hcp}} \geq 37$).

**Movie S1.**

Deformation microstructure evolution and dislocation patterning in grain of a single active slip plane. The hcp-coordinated atoms are colored in red. Shockley partial dislocations are green, and stair-rod and Hirth partial dislocations are purple and red, respectively.

Link
https://www.dropbox.com/s/3q2yv6haiqr6srv/Supplementary%20Video%201.mp4?dl=0

**Movie S2.**

Deformation microstructure evolution and dislocation patterning in the grain of double active slip planes.

Link
https://www.dropbox.com/s/34e6jwfyq6kh0y3/Supplementary%20Video%202.mp4?dl=0

**Movie S3.**

Deformation microstructure evolution and dislocation patterning in the grain of multiple active slip planes.

Link
https://www.dropbox.com/s/1i40vsc14g4xb51/Supplementary%20Video%203.mp4?dl=0